\documentstyle[preprint,aps,floats,psfig]{revtex}
\catcode`@=11
\def\references{%
\ifpreprintsty
\bigskip\bigskip
\hbox to\hsize{\hss\large \refname\hss}%
\else
\vskip24pt
\hrule width\hsize\relax
\vskip 1.6cm
\fi
\list{\@biblabel{\arabic{enumiv}}}%
{\labelwidth\WidestRefLabelThusFar  \labelsep4pt %
\leftmargin\labelwidth %
\advance\leftmargin\labelsep %
\ifdim\baselinestretch pt>1 pt %
\parsep  4pt\relax %
\else %
\parsep  0pt\relax %
\fi
\itemsep\parsep %
\usecounter{enumiv}%
\let\p@enumiv\@empty
\def\theenumiv{\arabic{enumiv}}%
}%
\let\newblock\relax %
\sloppy\clubpenalty4000\widowpenalty4000
\sfcode`\.=1000\relax
\ifpreprintsty\else\small\fi
}
\catcode`@=12
\begin{document}
\def\beq{\begin{equation}}
\def\eeq{\end{equation}}
\def\d{\delta}
\def\fourG{{{}^{(4)}G}}
\def\4R{{{}^{(4)}R}}
\def\H{{\cal H}}
\def\K{{\kappa}}
\def\mh{m_h^{}}
\def\vev#1{{\langle#1\rangle}}
\def\gev{{\rm GeV}}
\def\tev{{\rm TeV}}
\def\fbi{\rm fb^{-1}}
\def\lsim{\mathrel{\raise.3ex\hbox{$<$\kern-.75em\lower1ex\hbox{$\sim$}}}}
\def\gsim{\mathrel{\raise.3ex\hbox{$>$\kern-.75em\lower1ex\hbox{$\sim$}}}}
\newcommand{\hmu}{{\hat\mu}}
\newcommand{\hnu}{{\hat\nu}}
\newcommand{\hrho}{{\hat\rho}}
\newcommand{\hh}{{\hat{h}}}
\newcommand{\hg}{{\hat{g}}}
\newcommand{\hk}{{\hat\kappa}}
\newcommand{\tA}{{\widetilde{A}}}
\newcommand{\tP}{{\widetilde{P}}}
\newcommand{\tF}{{\widetilde{F}}}
\newcommand{\th}{{\widetilde{h}}}
\newcommand{\tp}{{\widetilde\phi}}
\newcommand{\tchi}{{\widetilde\chi}}
\newcommand{\te}{{\widetilde\eta}}
\newcommand{\vn}{{\vec{n}}}
\newcommand{\vm}{{\vec{m}}}
\newcommand{\A}{A}
\newcommand{\B}{B}
\newcommand{\mmu}{\mu}
\newcommand{\mnu}{\nu}
\newcommand{\ii}{i}
\newcommand{\jj}{j}
\newcommand{\jl}{[}
\newcommand{\jr}{]}
\newcommand{\ml}{\sharp}
\newcommand{\mr}{\sharp}
\renewcommand{\theenumi}{\roman{enumi}}
\renewcommand{\labelenumi}{(\theenumi)}
\newcommand{\da}{\dot{a}}
\newcommand{\db}{\dot{b}}
\newcommand{\dn}{\dot{n}}
\newcommand{\dda}{\ddot{a}}
\newcommand{\ddb}{\ddot{b}}
\newcommand{\ddn}{\ddot{n}}
\newcommand{\pa}{a^{\prime}}
\newcommand{\pb}{b^{\prime}}
\newcommand{\pn}{n^{\prime}}
\newcommand{\ppa}{a^{\prime \prime}}
\newcommand{\ppb}{b^{\prime \prime}}
\newcommand{\ppn}{n^{\prime \prime}}
\newcommand{\fda}{\frac{\da}{a}}
\newcommand{\fdb}{\frac{\db}{b}}
\newcommand{\fdn}{\frac{\dn}{n}}
\newcommand{\fdda}{\frac{\dda}{a}}
\newcommand{\fddb}{\frac{\ddb}{b}}
\newcommand{\fddn}{\frac{\ddn}{n}}
\newcommand{\fpa}{\frac{\pa}{a}}
\newcommand{\fpb}{\frac{\pb}{b}}
\newcommand{\fpn}{\frac{\pn}{n}}
\newcommand{\fppa}{\frac{\ppa}{a}}
\newcommand{\fppb}{\frac{\ppb}{b}}
\newcommand{\fppn}{\frac{\ppn}{n}}

\newcommand{\dA}{\dot{A_0}}
\newcommand{\dB}{\dot{B_0}}
\newcommand{\fdA}{\frac{\dA}{A_0}}
\newcommand{\fdB}{\frac{\dB}{B_0}}

\newcommand{ \slashchar }[1]{\setbox0=\hbox{$#1$}   
   \dimen0=\wd0                                     
   \setbox1=\hbox{/} \dimen1=\wd1                   
   \ifdim\dimen0>\dimen1                            
      \rlap{\hbox to \dimen0{\hfil/\hfil}}          
      #1                                            
   \else                                            
      \rlap{\hbox to \dimen1{\hfil$#1$\hfil}}       
      /                                             
   \fi}                                             %

\tighten
\preprint{ \vbox{
\hbox{MADPH--00-1184}
\hbox{hep-ph/0006275}}}
\draft
\title{ Cosmology and Hierarchy in Stabilized Warped 
Brane Models }
\author{V. Barger$^1$, T. Han$^1$, T. Li$^1$, J. D. Lykken$^{1,2}$ 
and D. Marfatia$^1$\footnote{
barger@oriole.physics.wisc.edu, than@pheno.physics.wisc.edu, 
li@pheno.physics.wisc.edu, \\ lykken@fnal.gov, 
marfatia@pheno.physics.wisc.edu }}
\vskip 0.3in
\address{$^1$Department of Physics, University of Wisconsin--Madison, WI 53706}

\vskip 0.15in
\address{$^2$Theory Group, Fermi National Accelerator Laboratory, 
Batavia, IL 60510 }
\vskip 0.15in
\vskip 0.1in

\maketitle

\begin{abstract}
{\rm
We examine the cosmology and hierarchy of scales 
in models with branes immersed in a
five-dimensional curved spacetime subject to radion stabilization.
When the radion field is time-independent and the inter-brane
spacing is stabilized, the universe can naturally find itself in the 
radiation-dominated epoch. This feature is independent of the 
form of the stabilizing potential. 
We recover the standard Friedmann equations without assuming 
a specific form for the bulk energy-momentum tensor.
In the models considered, if the observable brane has positive 
tension, a solution to the hierarchy problem requires 
the presence of a negative
tension brane somewhere in the bulk. We find that the
string scale can be as low as the electroweak scale.
In the situation of self-tuning branes 
where the bulk cosmological constant is set to zero, 
the brane tensions have hierarchical values.
In the case of a polynomial stabilizing potential
no new hierarchy is created.}
\end{abstract}

{{\bf I. Introduction.}}
It has recently been realized that the string scale
can be much lower than the Planck scale and even close to the
electroweak scale \cite{ewstring}. 
A low string scale provides new avenues on solving the 
hierarchy problem \cite{add,rs1}. The argument resides
in the fact that a low string scale ($M_X$) may result in
the apparent size of the Planck scale ($M_{Pl}$) due to the 
existence of a large volume ($R^n$) of compact extra 
dimensions, $ M_{Pl}^2 \sim M_X^{n+2} R^n $ \cite{add}.
Such a scenario may lead to a rich phenomenology at low energies
and is thus testable at collider experiments \cite{pheno}.

Recently, a model involving just one extra dimension with a
background $AdS_5$ metric was proposed by Randall and Sundrum \cite{rs1} (see also
Ref. \cite{merab}).
In this scenario, 
two branes (one with positive tension and
the other with negative tension) are located at the fixed points 
of an $S^1/\mathbf{Z_2}$ orbifold in a bulk 
with negative cosmological constant. An exponential hierarchy 
between the physical scales on the two branes is generated due to the curved
spacetime, providing an explanation for the large hierarchy between the 
weak and Planck scales. The model is amenable to a holographic 
interpretation motivated by string theories \cite{holography}. 
However, the Randall-Sundrum model has some drawbacks. 
First, a perfect fine-tuning among the brane tensions and the
bulk cosmological constant is needed to guarantee a 
static solution for the warped metric of spacetime.
Mechanisms for stabilizing the brane locations
via interactions between a bulk scalar field (called the radion) 
and the branes were suggested in \cite{stab} and elaborated in 
\cite{DFGK}, where an elegant solution that accounts for 
the back-reaction of the scalar profile on the geometry is outlined. 
An overall fine-tuning equivalent to setting the four-dimensional 
cosmological constant to zero is still present. 
Second, it was found that the brane world may not lead to the standard 
cosmology \cite{branecosmo,cosmology1}. 
In particular, the Hubble parameter
$H$ was found to be proportional to the matter density 
$\rho$ \cite{cosmology1}, in contradiction
with the usual $H\sim \sqrt{\rho}$ behavior. 
Although this can be remedied by a fine-tuned cancellation
between the brane tension and the bulk cosmological constant \cite{cosmology2},
only with a negative energy density on the observable brane 
is the standard cosmology recovered \cite{cosmology3}. 
Much attention has been devoted 
to studying cosmology without an explicit stabilization 
mechanism \cite{cosmology}.
The connection between radion stabilization and
cosmology was explored in \cite{cosmology3,cosmology4,trace}, 
and the standard cosmology can be obtained if the radion is time-dependent 
\cite{cosmology3}.
There have been attempts to modify the
Randall-Sundrum model so that the observable
brane has positive tension \cite{lr,ross,general,k}, since
the localization of matter and gauge  fields on positive tension 
branes is well understood in string theory.
An initial study with two positive tension branes 
was carried out \cite{lr}, incorporating localized
gravity in a noncompact $AdS_5$ geometry \cite{rs2}. 
The necessary hierarchy can be generated between the Planck
and electroweak scales by placing the hidden and observable branes 
at specific locations in the infinite dimension.
We will refer to this as the Lykken-Randall model. 

In this letter we examine the cosmology and hierarchy in models 
with radion stabilization. We shall adopt the formalism 
of Ref.~\cite{DFGK} to stabilize the brane separation. 
We call this the Solution Generating Technique. 
We generalize it to the case of branes
with arbitrary tensions and no relation between 
the metrics on either side of the branes. Using this technique we study
the cosmology resulting from radion stabilization and find a cancellation
between the bulk cosmological constant and the brane tension. This approach
is rather different in philosophy from the one adopted in
Refs.~\cite{cosmology4,trace,k}, where the bulk energy-momentum
tensor extracted from the linearized field equations 
is chosen specifically to get the conventional cosmology.
We obtain the important result 
that the stabilization of the inter-brane spacing can
naturally lead to the cosmology of the 
radiation-dominated universe. Specifically, this 
is a consequence of requiring
consistency in the equation of motion of the radion
before and after perturbing the solutions by placing  
matter energy density on the observable brane.
We argue that to obtain the complete evolution of our universe, 
a time-dependence of the radion field on the brane should be introduced. 
We speculate on the existence of some
new dynamics that causes the transition from one epoch to the next. 

Guided by cosmology, 
we explore the consequences on the hierarchy between the Planck and
 electroweak scales. 
As concrete examples, we study two classes of models. The first
is the ``self-tuning'' model, motivated by recent attempts to solve
 the cosmological constant 
problem \cite{cosmoprob}.  A dilaton-like
coupling of a bulk scalar field with a brane was shown to result 
in a vanishing 
bulk cosmological constant.  The result persists irrespective of 
the tension on the brane. This feature is referred to as self-tuning. 
The other model involves a 
Higgs-like radion potential, similar to that in \cite{stab}. 
In both models, the cosmology on the observable brane is independent of the 
configuration of branes and the potential that leads to
radion stabilization. All that is required is {\it{some}} 
stabilizing potential and that the observable brane has positive tension. 
To generate the hierarchy of scales, at least one hidden brane 
with negative tension is required.  The latter 
cannot be positioned at an orbifold fixed point.

The self-tuning brane model has the following properties:
\vspace{-2.5mm}
\begin{enumerate}
\addtolength{\itemsep}{-2.5mm}
\item{The model illustrates the unique minimal configuration from 
which the hierarchy of scales can be obtained without fine-tuning. 
There are two positive tension branes (one of which is the observable
brane), and one negative tension brane. 
The values of the brane tensions become hierarchical. }
\item{A dilatonic coupling between the branes and the bulk stabilizes the
inter-brane spacings. Thus the radion may be identified with the dilaton.}
\item{The same coupling ensures self-tuning of the branes to be 
flat and the bulk cosmological constant to be zero. 
The tree-level contribution to the four-dimensional 
cosmological constant is eliminated. 
We will truncate the space to avoid curvature singularities. }
\end{enumerate}
\vspace{-2.5mm}
The model with a Higgs-like radion potential possesses the following
properties:
\vspace{-2.5mm}
 \begin{enumerate}
\addtolength{\itemsep}{-2.5mm}
\item{The minimal set-up to generate the scale hierarchy requires one
 positive tension observable brane and one negative tension
hidden brane.}
\item{ The extra dimension is linearly infinite with finite proper volume.}
\item{The radion field is unbounded above and leads to the bulk 
cosmological constant being unbounded below.}
\end{enumerate}
\vspace{-2.5mm}
 
In Section II we present the Solution Generating Technique. Section III is
 devoted to studying the cosmology in a general setting. In Section IV we 
demonstrate a realization of a Lykken-Randall-like model 
with self-tuning branes. 
In Section V we perform a similar analysis but with a polynomial 
superpotential.  We conclude in Section VI.

{\bf {II. Solution Generating Technique.}}
It is possible to use a  gauged supergravity-inspired approach to reduce the 
nonlinear classical field equations of brane models with scalar-tensor 
gravity to a system of decoupled first order differential 
equations \cite{DFGK}.
Using the technique of \cite{DFGK}, the brane spacing can be stabilized. 
 We assume the scalar field to be 
static to prevent the four-dimensional Planck mass from being time-dependent.
The formalism is independent of whether the fifth dimension
is compact or noncompact. We shall
present the arguments for the case of a noncompact dimension.  

\begin{figure}[tb]
\centerline{\psfig{file=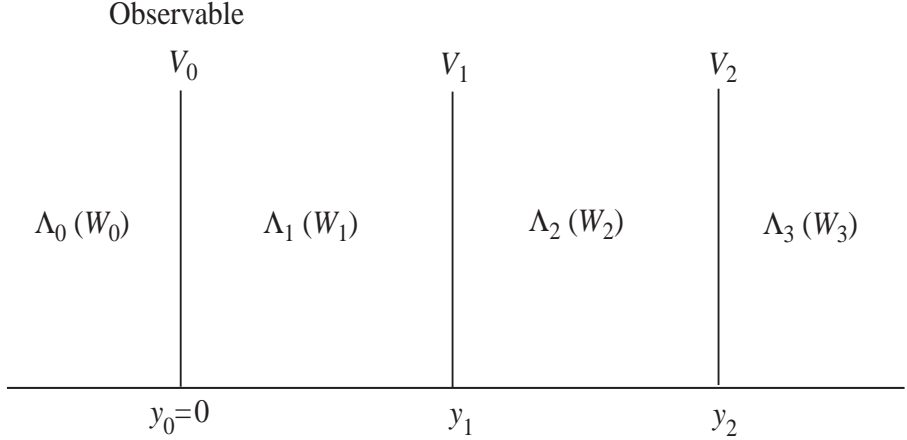,width=12cm,height=6cm}}
\bigskip
\caption[]{The configuration of branes in the bulk. 
$V_i$ is the tension of the brane at $y_i\,$, 
$\Lambda_i$ is the cosmological constant in the slice of $AdS_5$ 
between $y_{i-1}$ and $y_i$ and $W_i$ is the corresponding superpotential.  }
\label{config}
\end{figure}

We assume the presence of three 3-branes in the space $(-\infty,\infty)$, 
 at $y_0=0\,$, $y_1$ and $y_2$ with the observable brane
 located at $y_0\,$. We will refer to the branes at $y_1$ and $y_2$ as
 hidden branes.  The four-dimensional metric on
the brane labelled by its position $y_i$ is 
$g_{\mu \nu}^{(i)} (x^{\mu}) \equiv g_{\mu \nu}(x^{\mu}, y=y_i)\,$,
where $g_{AB}$ is the five-dimensional metric and $A,B=0,1,2,3,5$ and 
$\mu,\nu=0,1,2,3$. We use the metric signature $(-,+,+,+,+)$.
The five-dimensional gravitational action including a scalar field $\phi(y)$ is 
$S = S_{Gravity} + S_{Brane}$ with 
\begin{eqnarray}
S_{Gravity}&=& 
\int d^4 x  ~dy~ \sqrt{-g} \{ 
{1\over 2\, \kappa^2}\,  R -{1\over 4 \, \kappa^2}\,   \partial_A \phi\,  \partial^A \phi - \Lambda (\phi)  \} 
~,~\, \\
S_{Brane} &=& -\sum_{i=0}^{2} \int d^4 x  ~dy~  
\sqrt{-g^{(i)}}\:V_{i}( \phi)\,
 \delta(y-y_i)
~,~\,
\end{eqnarray}
where $\kappa^2=8\,\pi\,G_N^{(5)}=M_X^{-3}$ is the five-dimensional coupling constant of gravity, $M_X$ is the Planck scale in five dimensions, 
$R$ is the curvature scalar and $V_i(\phi_i)$ is the tension of the brane
at $y_i$. $\Lambda(\phi)$ is the potential of the field $\phi$ in the bulk
and is interpreted as the cosmological constant although it has a $\phi$-dependence.  
We allow it to be discontinuous at the branes, but continuous in each section.
 We write $\Lambda(\phi)$ as $\Lambda_0(\phi)\,$ if $\,y<0\,$, $\Lambda_1(\phi)\,$ if $\,0<y<y_1\,$ as $\Lambda_2(\phi)\,$ if $\,y_1<y<y_2\,$ and 
$\Lambda_3(\phi)\,$ if $\,y>y_2\,$ 
(see Fig.~\ref{config}).
%

The five-dimensional Einstein equations arising from the above 
action are
\begin{eqnarray} 
G_{AB}\equiv R_{AB}-{1 \over 2 } g_{AB} R=\kappa^2\,T_{AB} &=& 
{1 \over 2 }\, ( \partial_A \phi\,  \partial_B \phi
-{1 \over 2 }\, g_{AB}\, ( \partial \phi)^2)-
\kappa^2\, g_{AB} \Lambda (\phi)  
\nonumber\\&& -\kappa^2\, \sum_{i=0}^{2}
V_i (\phi) \sqrt{{g^{(i)} \over g}} ~g_{\mu \nu}^{(i)} 
~\delta^\mu_A \delta^\nu_B ~\delta(y-y_i)\,,
\label{emtensor}
\end{eqnarray}
where $R_{AB}$ is the five-dimensional Ricci tensor. The most general 
five-dimensional metric that respects four-dimensional 
Poincar\'{e} symmetry is
\begin{equation} 
ds^2 = e^{2\,  A(y)}\ \eta_{\mu \nu}\ dx^{\mu} dx^{\nu}
 + (dy)^2 ~.~\, 
\end{equation} 
The factor $e^{2\,  A(y)}$ is commonly called a ``warp factor''. 
The equation of motion of $\phi$ is
\begin{eqnarray} 
 \phi'' + 4\, A'\, \phi' &=& 2\, \kappa^2\left(   {\partial \Lambda(\phi) 
\over \partial \phi}  
+ \sum_{i=0}^2 
{\partial V_i(\phi) \over \partial\phi}~\delta(y-y_i)\right)  \,,
\label{motion1}
\end{eqnarray}
and the Einstein equations can  be written as
\begin{equation} 
   A'' = -{1 \over 6} \phi'^2 - {\kappa^2 \over 3} \sum_{i=0}^2 
 V_i(\phi) ~\delta(y-y_i)   \,,\ \ \ \ \ \ \  
   A'^2 =  {1 \over 24} \phi'^2 -{\kappa^2 \over 6} \Lambda(\phi)  \ .
\label{motion3}
\end{equation}
Here a prime denotes a derivative with respect to $y$. 
The jumps corresponding to the presence of the branes are
\begin{equation} 
A'\Big|^{y_i+\epsilon}_{y_i-\epsilon} = 
- {\kappa^2 \over 3} V_i(\phi_i)\,,  \ \ \ \ \ \ 
\phi'\Big|^{y_i+\epsilon}_{y_i-\epsilon} = 
2\,\kappa^2  {\partial V_i(\phi) \over \partial\phi}
\Bigg|_{\phi=\phi_i}\,,
\label{bc}
\end{equation}   
where $\phi_i \equiv \phi(y_i)$.  

Let $W(\phi)$ be
 any sectionally continuous
function (which we call the superpotential), with sectional functions $W_{i}(\phi)$ 
defined analogous to $\Lambda_{i}(\phi)$ (see Fig.~\ref{config}).
 Taking
\begin{eqnarray} 
2\,\kappa^2\,\Lambda(\phi)&=& {1 \over 2}\,\left( \partial W(\phi) 
\over \partial \phi \right)^2
- {1 \over 3}\, W(\phi)^2 \, ,
\label{cosmoconst}
\end{eqnarray}
it is possible to show that a solution to the equations,
\begin{equation} 
\phi' = {\partial W(\phi) \over \partial\phi} \,,\ \ \ \ \ \ 
\label{phi'} 
   A' = -{1 \over 6}\, W(\phi) \,,
\end{equation}
subject to the constraints,
\begin{equation} 
W(\phi)\Big|^{y_i+\epsilon}_{y_i-\epsilon} = 
 {2\,\kappa^2} V_i(\phi_i)\,,  \ \ \ \ \ \ 
{\partial W(\phi) \over \partial\phi}\Bigg|^{y_i+\epsilon}_{y_i-\epsilon} = 
 {2\,\kappa^2}  {\partial V_i(\phi) \over \partial\phi}
\Bigg|_{\phi=\phi_i}
\end{equation}  
is also a solution to the system of equations ($\ref{motion1}-\ref{bc}$). 
By solving Eq.~(\ref{phi'})
in the bulk and applying boundary conditions on the branes, 
we can determine the 
 locations of the branes and hence their separation.
 Note that up to an arbitrary function $\sum_{n=2}^\infty \gamma_n\,(\phi-\phi_i)^n\,$, 
the brane tension is 
completely determined by the superpotential and the value of $\phi$ on the
brane, 
\begin{equation} 
 {2\,\kappa^2}\, V_i(\phi) =W_{i+1}(\phi_i)-W_i(\phi_i)+\left({\partial \over \partial\phi} (W_{i+1}(\phi)-W_{i}(\phi))\right)
\Bigg|_{\phi=\phi_i}\,(\phi-\phi_i)\,.
\label{tension}
\end{equation}
 The solution involves fine-tuning even though it appears not to be the case.
 There are six constraints
arising from the jump conditions on the three branes, but only five
integration constants; the 
equations of motion and the jumps depend only upon $A'(y)$ and $A''(y)$ thereby
rendering the value of $A$ on one of the branes irrelevant \cite{DFGK}.

{\bf {III. Cosmology: General Considerations.}}
 Our starting point is the most general five-dimensional metric that
preserves three-dimensional rotational and translational invariance. Thus, we
adopt the cosmological principle of isotropy and homogeneity on the observable
brane. However, $y$-dependence is maintained in the metric tensor since 
isotropy
is broken in the fifth dimension due to the presence of branes. We
consider the metric to be of the form
\begin{equation}
ds^{2}=-n^{2}(\tau,y) d\tau^{2}+a^{2}(\tau,y)d{\mathbf{x}^2}+b^{2}
(\tau,y)dy^{2}.
\label{metric}
\end{equation}
The Einstein tensor for this metric is given by \cite{cosmology1},
\begin{eqnarray}
G_{00} &=& 3\left\{ \fda \left( \fda+ \fdb \right) - \frac{n^2}{b^2} 
\left(\fppa + \fpa \left( \fpa - \fpb \right) \right)  \right\}, 
\label{ein00} \\
 G_{\ii\jj} &=& 
\frac{a^2}{b^2} \delta_{ij}\left\{\fpa
\left(\fpa+2\fpn\right)-\fpb\left(\fpn+2\fpa\right)
+2\fppa+\fppn\right\} 
\nonumber \\
& &+\frac{a^2}{n^2} \delta_{ij} \left\{ \fda \left(-\fda+2\fdn\right)-2\fdda
+ \fdb \left(-2\fda + \fdn \right) - \fddb \right\},
\label{einij} \\
G_{05} &=&  3\left(\fpn \fda + \fpa \fdb - \frac{\dot{a}^{\prime}}{a}
 \right),
\label{ein05} \\
G_{55} &=& 3\left\{ \fpa \left(\fpa+\fpn \right) - \frac{b^2}{n^2} 
\left(\fda \left(\fda-\fdn \right) + \fdda\right) \right\}.
\label{ein55} 
\end{eqnarray} 
where a dot denotes a derivative with respect to $\tau$.
Note that the time-independent solution of the previous section corresponds to 
$a(\tau,y)=n(\tau,y)=e^{A(y)}$ and  \mbox{$b(\tau,y)=1$}. 
 We will maintain the assumption that the stabilizing potential is static,
$\dot{b}=0\,$, and without loss of generality we set $b=1$. 
The energy-momentum tensor can be decomposed into two parts: 
a contribution from fields on the observable brane, 
\mbox{${\tilde{T}^A}_{\ \ B}=diag(-\rho_0,p_0,p_0,p_0,0)\,\delta(y)\,$}, 
and the 
contribution ${\check{T}^A}_{\ \ B}$ of all other sources, {\it i.e.} 
bulk fields and matter on the other brane:
\mbox{${T^A}_B={\tilde{T}^A}_{\ \ B}+{\check{T}^A}_{\ \ B}\,$}. 
In the time-independent case, 
\mbox{${\tilde{T}^A}_{\ \ B}=diag(-V_0,-V_0,-V_0,-V_0,0)\,
\delta(y)\,$}.
The jump conditions on the observable brane are \cite{cosmology1}
\begin{equation}
{{\jl a^\prime \jr} \over {a_0}}=-{{\K^2} \over 3} \rho_0\, \ \ \  {\rm{and}}\ \ \  
{{\jl n^\prime \jr} \over {n_0}}= {{\K^2} \over 3} \left( 3\,p_0 + 2\,\rho_0 
\right), 
\label{nnrho}
\end{equation}
where $[a^\prime] =a^\prime(+\epsilon)-a^\prime(-\epsilon)\,$,
and functions with the subscript $0$ are evaluated 
on the observable brane.
 On taking the jump of the (0,5) component of
Einstein's equations, 
these conditions lead to 
the energy conservation equation, 
\begin{equation}
\dot{\rho_0} + 3(p_0+\rho_0) {{\dot{a_0}}\over {a_0}}  = 0\,.
\end{equation}
which is independent of $\dot{b}$.
 Taking the jump of the (5,5) component of
 Einstein's equations, we get
 \begin{equation}
3\,p_0\, {{\vev{a^\prime }}\over {a_0}} = \rho_0 
 {{\vev {n^\prime }}\over {n_0}}+{[\check{T}_{55}] \over a_0}\,, \label{mean}
\end{equation}
where 
\begin{equation}
2\,\vev{a'}=a'(+\epsilon)+a'(-\epsilon)\,.
\label{avg}
\end{equation}
We choose $n_0=1$, which amounts to identifying $\tau$ with time in 
conventional cosmology. Evaluating the (5,5) component of
 Einstein's equations on either side of the observable brane and adding, we 
obtain
\begin{eqnarray}
\left({\dot{a_0} \over a_0}\right)^2 + {\ddot{a_0} \over a_0}= -{\kappa^4 \over 36}\,\rho_0\,(\rho_0+3\,p_0)
-{\kappa^2 \over 3} \,\vev{\check{T}_{55}}
+{{\vev{a'}}^2 \over a_0^2} (1+{3\,p_0 \over \rho_0})-{\vev{a'}\,[\check{T}_{55}] \over \rho_0\, a_0^2} \,.
\end{eqnarray}
Usually, motivated by Ho\v{r}ava-Witten supergravity 
\cite{Horava}, a $\mathbf{Z_2}$ symmetry is imposed on the solutions. 
We simplify the above result by requiring the solutions to Einstein's 
equations to obey a ``$Z_2$ symmetry'' in the neighborhood of  
the observable brane. By this we simply mean 
\begin{equation}
W(\phi(+\epsilon))=-W(\phi(-\epsilon))\,,
\label{Z2}
\end{equation} 
because it leads to the warp factor being symmetric on either side of 
the observable brane. 
By a contextual abuse of terminology, we will call this a 
``local $Z_2$ symmetry''. 
(Of course, in no sense are we gauging the symmetry). 
To create a distinction, we will reserve the bold font for the ``global'' 
 $\mathbf{Z_2}$ symmetry. 
We are therefore left 
with the following Friedmann-like equation:
\begin{equation}
\left({\dot{a_0} \over a_0}\right)^2 + {\ddot{a_0} \over a_0}= -{\kappa^4 \over 36}\,\rho_0\,(\rho_0+3\,p_0)
-{\kappa^2 \over 3} \,\vev{\check{T}_{55}}
 \,,
\label{FRW} 
\end{equation}
derived in \cite{cosmology1}. Let us emphasize the 
two assumptions on which our results will hinge. 
They  are:
\vspace{-2.5mm}
\begin{enumerate}
\addtolength{\itemsep}{-2.5mm}
\item{ The extra dimension is assumed to be stable before studying cosmology.} 
\item{ The solutions satisfy a $Z_2$ symmetry in the immediate neighborhood of
the observable brane, Eq.~(\ref{Z2}).}
\end{enumerate}
\vspace{-2.5mm}
In finding the static solution of the previous section, we 
ignored the matter energy densities on the branes by assuming that they 
are negligible in comparison to the brane tensions. 
We now include their contribution as a perturbation
to the ``matter-less'' solution. 
Thus, we can study the resulting cosmology by making the ansatz
\begin{equation}  
 \rho_0=\rho+V_0\,,\ \ \ \  p_0=p-V_0\,,
\label{pert}
\end{equation}
and Eq.~(\ref{FRW}) becomes
\begin{eqnarray}
\left({\dot{a_0} \over a_0}\right)^2 + {\ddot{a_0} \over a_0}= 
{\kappa^4\over 18}\,V_0^2-\,{1 \over 18}\,\vev{W(\phi_0)^2} 
+{\kappa^4\over 36}\,
V_0\,(\rho-3\,p)-{\kappa^4\over 36}\,\rho\,(\rho+3\,p)\,.
\label{FRW1.5}
\end{eqnarray}
Here we have used
\begin{equation}
\kappa^2\,\vev{\check{T}_{55}}\,={1 \over 6}\,\vev{W(\phi_0)^2}\,,
\end{equation}
which is obtained by inserting (\ref{cosmoconst}) and (\ref{phi'}) 
into (\ref{emtensor}).
Notice that $W(\phi_0)$ is proportional to $A'(0)$ and is therefore 
not well-defined. However, $\vev{W(\phi_0)^2}$ is well-defined. 
On account of the local $Z_2$ symmetry 
 and Eq.~(\ref{tension}), we have
\begin{equation}
W(\phi(+\epsilon))^2=W(\phi(-\epsilon))^2= \kappa^4\, V_0^2\,.
\end{equation}
The definition of $\vev{W(\phi_0)^2}$ is analogous to Eq.~(\ref{avg}), and we obtain
\begin{equation}
\vev{W(\phi_0)^2}= \kappa^4\,V_0^2\,.
\end{equation}
This result relies heavily on the existence of the local $Z_2$ symmetry. 
The first two terms on the right hand side of Eq.~(\ref{FRW1.5}) cancel out and we are 
left with  
\begin{equation}
\left({\dot{a_0} \over a_0}\right)^2 + {\ddot{a_0} \over a_0}= 
{\kappa^4\over 36}\,V_0\,(\rho-3\,p)-{\kappa^4\over 36}\,\rho\,(\rho+3\,p)\,.
\label{FRW2}
\end{equation}
The leading term on the right-hand side reproduces the
standard cosmology if we make the identification, $\kappa^4V_0=6/M_{Pl}^2$.
It is essential for the observable brane to 
possess positive tension to arrive at the correct 
Friedmann equations in spite of an 
explicit radion stabilization. Furthermore,
 a specific form of the bulk energy-momentum 
tensor was not chosen to implement the cancellation. 
Beyond the conventional fine-tuning one does not need additional machinery 
to obtain the usual cosmology. 

In introducing the perturbation 
(\ref{pert}), it is
no longer obvious that the solutions remain consistent. Let us consider 
the equation of motion of $\phi$. On the observable brane it is 
\begin{equation}
\phi''+ \left(3\, \fpa+\fpn \right)\,\phi'=2\,\kappa^2 
{\partial \over \partial\phi} (\Lambda(\phi)+V_0(\phi))\,,
\end{equation}
where all quantities are evaluated at $y_0$. 
 We started with just tension on the branes and demanded that
the radion stabilize the configuration of branes. Having obtained this static 
solution we then proceeded to consider the effect of matter on the 
observable brane.
Let us require that the 
positions of the branes be unchanged by appealing to the
stability of such a scenario. 
This is equivalent to saying that $\phi$ and $\Lambda(\phi)$ 
are unchanged before and after the introduction of the matter 
energy density. Then consistency requires,
\begin{equation}
 \left(3\, \fpa+\fpn \right)\Big|_{0\,,\,Static}=\left(3\, 
\fpa+\fpn \right)\Big|_{0\,,\,Perturbed}\ \,. 
\end{equation}
Again, using the local $Z_2$ symmetry and the jump conditions 
(\ref{nnrho}), we get
\begin{equation}
-\,{\kappa^2 \over 2}\, V_0-{\kappa^2 \over 6}\, V_0=
-\,{\kappa^2 \over 2}\,(\rho+V_0)+
{\kappa^2 \over 6}\,(3\,p+2\,\rho-V_0)\,,
\end{equation}
which leads to the condition for a radiation-dominated (RD) universe,
\begin{equation}
\rho=3\,p\,.
\label{RD}
\end{equation}
The interpretation of the above constraint is interesting.  When matter
on the observable brane is radiation, the inter-brane spacing is 
identical to the case when there is no matter on the brane. Conversely,
when the brane location is unaffected by the matter-perturbation, 
the universe is RD. This observation is consistent with the
fact  that the radion couples to the 
trace of the energy-momentum tensor \cite{cosmology3,trace}, 
which is zero for radiation.
It may be possible to identify the process of radion stabilization 
with inflation and reheating 
and the time at which the inter-brane spacing 
becomes stable marks the end of reheating.
The RD universe then ensues.

 To study cosmology at
lower temperatures, we need the radion to be time-dependent. 
The radion can be written as 
\begin{equation}
\phi(\vec{x},t,y)=\phi(\vec{x},t)\,\phi(y)\,,
\end{equation}
the form of which encodes the requirement that the bulk 
remain static (so as to maintain a non-fluctuating Planck scale). 
In our previous analysis we have set $\phi(\vec{x},t)=1$. 
The Solution Generating Technique is still 
applicable provided the derivatives of $\phi(\vec{x},t)$ with respect 
to $\vec{x}$ and $t$ are negligible. 
It is conceivable that $\phi(\vec{x},t)$ plays a role in the evolution from 
a RD universe to a matter-dominated universe and is perhaps responsible for 
the transition to an accelerating universe as in quintessence models \cite{k-essence}.

{\bf {IV. Self-tuning Flat Branes.}}
In this section we study the case where the superpotential takes on the form
of the tree-level dilaton coupling. This will lead us to the case of a 
vanishing bulk cosmological constant \cite{cosmoprob}. 
The resulting $\phi$ has two
singularities at finite distances on either side of the observable brane. 
We will describe how these singularities can be dealt with. 

Consider a superpotential of the exponential form
\begin{eqnarray}
W_{i}(\phi) &=& \omega_{i}\, e^{-\beta \phi}\,,
\end{eqnarray}
for which
\begin{equation}
12\,\kappa^2\,\Lambda_{i}(\phi)= (3\, \beta^2-2)\, 
\omega_{i}^2\, e^{-2 \beta \phi}\,.
\end{equation} 
For $\beta^2=2/3$, we have the important result that 
$\Lambda=0$ \cite{cosmoprob}. Henceforth,
we restrict ourselves to this choice. When we have not committed to the sign
of $\beta$, we will leave it explicit in the equations. With  $\beta^2=2/3$
 the branes are flat and will remain so, independent of the matter on 
them (hence the expression ``self-tuning flat branes''). 

As we pointed out in Sec.~II, the tension on the branes is fixed by the
superpotential. As long as the tension satisfies Eq.~(\ref{tension}), 
it is irrelevant what functional form it takes, {\it i.e.} 
the particular form of the tension we choose 
is simply a calculational device with no bearing on the physics. 
We take
\begin{eqnarray}
 {12\,\kappa^2}\,V_i(\phi) &=& (\omega_{i+1}-\omega_{i})\, e^{-\beta \phi}\,.
\label{t}
\end{eqnarray}
Then it can be shown that
\begin{equation}
\phi(y)={1 \over \beta} \ln[{- ( \sum_{i=0}^2 k_i\, |y-y_i| + k_c\, y) + c}]
\,,\ \ \ \ \  \label{ph}
A(y)={1 \over {6\, \beta}} \phi(y)+h\,,
\end{equation}
where
$k_i=  (\omega_{i+1}-\omega_i)/ 3\,$ and
$k_c=  (\omega_{0}+\omega_{3})/ 3\,$.
Here $c$ is a constant that can be fixed by the boundary 
conditions on any brane. 
Conventionally, the constant $h$ is set to zero, but its value does not
affect the hierarchy of scales. Let us call the argument of the logarithm in Eq.~(\ref{ph}), $f(y)$. Then,
\begin{equation}
f(y) \equiv e^{\beta \phi(y)} = -{2\over 3}\,\omega_i\,y+c_i\ \ \   {\rm{for}}
\ \ \ y_{i-1}\leq y \leq y_i\,,
\label{f}
\end{equation}
where $c_i$ is a constant of integration. Note that when 
$\omega_i$ is positive (negative), $e^{\beta \phi}$ falls (rises) linearly.
 When $f$ vanishes, $\phi$ diverges and the warp factor vanishes. 
 The space collapses to a point at these locations and these points 
can be identified with horizons.
  Assume that the horizons at $y_{min}$ 
and $y_{max}$ are such that $y_{min}<0$ and $y_{max}>y_{2}$.  
We will truncate the space at the horizons.
There is much debate about the
justification of this procedure if one hopes to solve the cosmological
constant problem. It has been claimed that
the four-dimensional cosmological constant vanishes only when the singularities
contribute to the vacuum energy \cite{Nilles}. If negative tension branes are introduced at the 
singularities, it is possible to set the four-dimensional cosmological 
constant to zero, but not without fine-tuning \cite{Nilles}. 
Attempts have been made to
find bulk potentials such that self-tuning remains while simultaneously 
removing the singularities. It has been found \cite{st} that if the 
singularities are removed, gravity is no longer localized and the 
four-dimensional Planck scale diverges. 
 We will
content ourselves with having $\Lambda=0$ as an improvement to the 
cosmological constant problem. We assume the presence of some
dynamics at the singularities that does not affect the global properties of
the solution and may resolve the problem of fine-tuning.
So that
$\phi$ be well-defined and accommodate our assumptions, 
we must impose the constraint,
\begin{equation} 
\sum_{i=0}^2  k_i >  |k_c|\  \Rightarrow   \begin{array}{ll}
\ \omega_3 >0\ \ {\rm{iff}}\ \  k_c <0\,, \ \ \ \ 
\ \omega_0 <0\ \ {\rm{iff}}\ \  k_c >0 \,.
\end{array} 
\label{constr}
\end{equation}
Now we can calculate the four-dimensional Planck scale in terms of the 
five-dimensional Planck scale $M_X$,
\begin{equation}
M_{Pl}^2 =  M_X^3\,\int e^{2\,A(y)}\, dy 
\end{equation}
to be
\begin{equation}
M_{Pl}^2 =  M_X^3\,{e^{2\, h}}\,\left[(
{1 \over \omega_1}-{1 \over \omega_0})\,e^{\phi_0/\beta}+
 ({1 \over \omega_2}-{1 \over \omega_1})\,e^{\phi_1/\beta}+
 ({1 \over \omega_3}-{1 \over \omega_2})\,e^{\phi_2/\beta}\right]\,.
\label{eq}
\end{equation}
We recall that the electroweak scale ($M_{EW}$) 
can be generated from the five-dimensional Planck scale $M_X$
via the square root of the warp factor \cite{rs1},
\begin{equation}
M_{EW} \simeq M_X\,e^{A(0)}=M_X\,e^{\phi_0/ (6\,\beta)}\, .
\end{equation}

{\it \underline{Two branes geometry:}}\ 
Let us specialize to the case  of just two branes. 
The formulae for the case of three branes apply by 
simply dropping the terms corresponding to the extra indices.
We choose
\begin{equation}
W_1(\phi)=-W_0(\phi)=\omega_1\, e^{-\beta \phi}\,,\ \ \ 
 W_2(\phi) =\omega_2\, e^{-\beta \phi}
\,.
\end{equation}
As required by cosmology,  $W(\phi)^2$ has a local $Z_2$ symmetry about 
the observable brane.  From 
Eq.~(\ref{t}), we need to impose $\omega_1>0$
  so that the observable brane has positive tension. 
 The four-dimensional Planck scale is
\begin{equation}
M_{Pl}^2 =  M_X^3 \left[
{2 \over \omega_1}\,e^{\phi_0/\beta}+({1 \over \omega_2}-{1\ 
\over \omega_1})\,e^{\phi_1/\beta} \right]\,.
\label{sup} 
\end{equation}
From Eq.~(\ref{f}), one can readily see that 
$e^{\phi_0/\beta}>e^{\phi_1/\beta}$. Since $\omega_0=-\omega_1<0\,$, 
the constraint
from Eq.~(\ref{constr}) requires $\omega_2>\omega_1\,$. This implies
that the hidden brane has positive tension. Therefore, the second term in
Eq.~(\ref{sup}) makes a negative contribution to $M_{Pl}^2$.  
The first term must be the dominant contribution for $M_{Pl}^2>0$
and $\phi_0/\beta\gg\phi_1/\beta$ must
be satisfied, where ``$\gg$'' implies a hierarchy 
of at most two orders of magnitude.
We would like the fundamental parameters $M_X$ and $\omega_i$
to be roughly of the same order of magnitude to avoid a fine-tuned
hierarchy of scales. We obtain
\begin{equation}
M_{Pl}\simeq M_X\,e^{\phi_0/ (2\,\beta)}\,,\quad 
M_{Pl}/M_{EW}\simeq e^{\phi_0/ (3\,\beta)}. 
\end{equation}
We need $\phi_0/(3\beta)\simeq 37$ to get
the correct hierarchy with $M_{Pl}\simeq 10^{19}$ GeV
and \mbox{$M_{EW}\simeq 10^3$ GeV}. 
This leads to \mbox{$M_X\simeq 10^{-5}$ GeV}. 
Interpreting $M_X$ as the string scale is now impossible.
The difficulty arises because both the Planck and electroweak 
scales are determined
by the value of $\phi$ on the observable brane. 
Any self-tuning brane model with only 
two branes shares this problem; $\phi$ will always 
be required to have its maximum value on the observable brane because 
of the local $Z_2$ symmetry.  

\begin{figure}[tb]
\centerline{\psfig{file=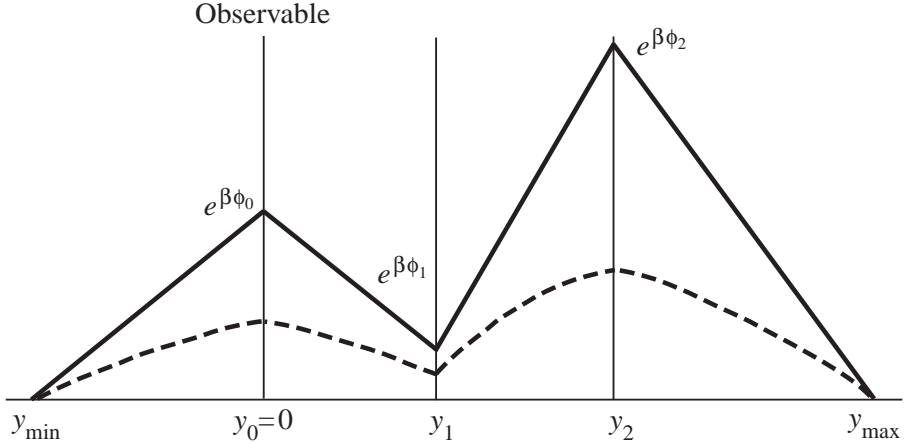,width=12cm,height=6cm}}
\bigskip
\caption[]{The profile of $e^{\beta \phi}$ (solid) and 
the warp factor (dashed) in the case of self-tuning branes. 
 In regions where $\omega_i$
is positive (negative), $e^{\beta \phi}$ falls (rises) linearly.
The brane at $y_1$ must have negative tension so that 
$e^{\beta \phi_2} > e^{\beta \phi_0}$.}
\label{self}
\end{figure}

{\it \underline{Three branes geometry:}}\ 
We construct a model with three branes 
in which the Planck scale will be generated by the value 
of $\phi$ on a neighboring brane. We investigate the superpotential
\begin{equation}
W_1(\phi)=-W_0(\phi)=\omega_1\, e^{-\beta \phi}\,,\ \ \ 
W_2(\phi) =\omega_2\, e^{-\beta \phi}
\,,\ \ \ W_3(\phi)=\omega_3\, e^{-\beta \phi}\,.
\end{equation}
Notice that $W(\phi)^2$ has a local $Z_2$ symmetry. 
We leave the sign of the tensions of the hidden branes 
unspecified for the time being. The Planck scale is given by
Eq.~(\ref{eq}) with $\omega_0=-\omega_1$.
Again, from Eq.~(\ref{f}), $e^{\phi_0/\beta}>e^{\phi_1/\beta}$. The
only way of getting $e^{\phi_2/\beta}>e^{\phi_0/\beta}$ is by choosing
 $\omega_2<0$. From Eq.~(\ref{t}) we can see that the brane at $y_1$ 
must have negative tension. Since there are no more branes in the bulk,
so that $e^{\phi/\beta}\to 0$ at $y_{max}$, we must have $\omega_3>0$.
Thus, the brane at $y_2$ has positive tension. 
This situation calls for two hidden
 branes, one with positive and the other with negative tension in a unique
configuration. The configuration of branes, 
the profiles of $e^{\beta \phi}$ and the warp factor 
$e^{2\,A(y)}=\sqrt{e^{\beta \phi}}$ are shown in Fig.~\ref{self}.
The model resembles the ``$+-+$'' model of Ref.~\cite{ross}, which however
is not derived by imposing constraints from radion stabilization and 
has a compactification on $S^1/{\mathbf{Z_2}}$ .
If we assume $M_X$ and $\omega_i$
to be of the same order of magnitude and
\mbox{$\phi_2/\beta \gg \phi_1/\beta\,,\ \phi_0/\beta$}, then
\begin{equation}
M_{Pl}\simeq M_X\,e^{\phi_2/ (2\,\beta)}\,, \quad
M_{Pl}/M_{EW}\simeq e^{{(\phi_2 -\phi_0/3)} /(2\,\beta)}\,.
\end{equation}
To obtain the correct hierarchy we must have 
$(\phi_2-\phi_0/3)/(2\beta) \simeq 37$.
By choosing appropriate values of $\phi_0$ and $\phi_2$, 
we are able to generate the hierarchy between $M_{Pl}$
and $M_{EW}$ for essentially any value of $M_X$ in between.
For illustration, we present two particularly interesting examples.
First, we can achieve $M_X \simeq M_{Pl}$ by taking
$\phi_2\to 0$ (or any other value that yields 
$e^{\phi_2}\simeq {\cal O}(1)$), 
corresponding to $\phi_0/\beta\simeq -220$.
At the other extreme, we can obtain $M_X \simeq M_{EW}$ by taking
$\phi_0\to 0$, corresponding to $\phi_2/\beta\simeq 75$.
We mention in passing that the solution that leads to 
$M_X \simeq M_{EW}$ is slightly less fine-tuned in terms of the
difference in $|\phi_i|$ than the one leading to 
the high energy string scale. A lighter string scale is preferred
in this sense. 

A couple of points are noteworthy. The solution presented represents 
the unique minimal configuration that allows for the generation of the 
hierarchy of scales without fine-tuning. 
The negative tension brane must lie between 
two positive tension branes.
In the case at hand, it is not 
possible to place the negative tension brane 
at the fixed point of an orbifold. 
The only possible discrete symmetry that can be 
imposed on $R^1$ is $\mathbf{Z_2}$. If we considered the orbifold  
$R^1/\mathbf{Z_2}$ with a fixed point at $y_1$, we would not be able to 
satisfy $e^{\phi_2/\beta}>e^{\phi_0/\beta}$. 
Since the negative tension brane is not at an orbifold fixed point,
the radion may have a problem with positivity of energy \cite{orbifold}. This
is an unpleasant circumstance but nevertheless, we assume
the model to be theoretically feasible. 
More problematic is the introduction of a 
new hierarchy problem. By inspecting Eq.~(\ref{t}) 
it can be seen that due to the
exponential dependence of the brane tensions on $\phi\,$, 
a large hierarchy is generated between the values of the 
tensions for even moderately different values of
$\phi$.

\begin{figure}[tb]
\centerline{\psfig{file=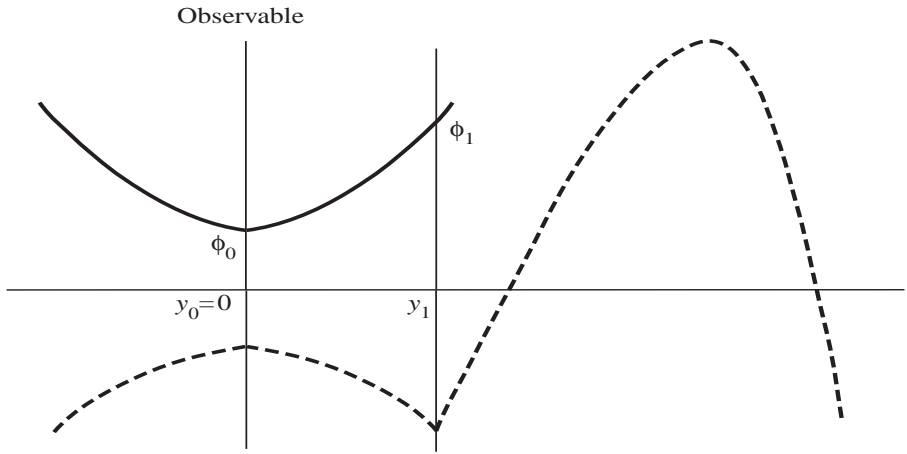,width=12cm,height=6cm}}
\bigskip
\caption[]{ Representative configurations of the radion (solid) and
 $A(y)$ (dashed) in the case of a quadratic superpotential. 
To generate the 
appropriate hierarchy of scales, the hidden brane at $y_1$
is required to have negative tension.}
\label{infinite}
\end{figure}

{\bf {V. Polynomial Superpotentials.}}
Here we consider the type of superpotential that leads to the 
stabilization mechanism
suggested in \cite{stab}. In \cite{DFGK} it was demonstrated that 
a quadratic superpotential results in the mechanism of \cite{stab}.
We have studied all geometries with two positive tension branes 
(with bounded and unbounded $\phi$)
and numerically scanned the parameter space. As in the model of the
previous section, we find that it is 
not possible to generate the appropriate scale hierarchy with only positive 
tension branes. 
We therefore study a model with two branes where the hidden brane has
 negative tension.
The first column of Table \ref{table} 
shows our particular choice of the polynomial superpotential,
which is guided by cosmology discussions with a $Z_2$ symmetry.
For simplicity, we have multiplied $y$ by $M_X$ to make it dimensionless.  
The second and third columns of Table \ref{table} present the
static solution to Einstein's equations, where $a_0$ is an irrelevant
integration constant which we set to zero. 
We find it necessary for the radion to be unbounded for $y>y_1$. 
This leads to $\Lambda(\phi)$ being unbounded below.
This is often seen in $AdS$ supergravity and offers no threat 
to the model \cite{DFGK}.
In the region $y<0$, the radion may or may not be bounded 
without affecting the hierarchy. We will choose it to be
unbounded in both regions, thus making the value of $\phi_0$ 
a global minimum. Figure \ref{infinite} 
illustrates the profiles of $\phi$ and $A(y)$  in the bulk. 
\begin{table}[tb]
\begin{center}
\begin{tabular}{|c|c|c|c|}
$ W(\phi)/M_X$ & $\phi(y)$ & $A(y)$ & Region   \\
\hline
$(\eta-\phi^2) $ & $\phi_0\,e^{2\,|y|} $  & $a_0+{1 \over 6}\,\eta\,
|y|-{\phi_0^2 \over 24}\,e^{4\,|y|} $  & $ y<0  $ \\
\hline
$-(\eta-\phi^2) $ & $\phi_0\,e^{2\,y} $  & $ a_0+{1 \over 6}\,
\eta\,y-{\phi_0^2 \over 24}\,e^{4\,y} $  
& $ 0<y<y_1     $ \\
\hline
$-(\xi-\phi^2) $ & $\phi_0\,e^{2\,y} $  & $ a_0+{1 \over 6}\,
\eta\,y_1+{1 \over 6}\,\xi\,(y-y_1)-{\phi_0^2 \over 24}\,e^{4\,y} $  
& $ y_1<y     $ \\
\end{tabular}
\caption[]{\small The solution to Einstein's equations 
in a model with a polynomial superpotential. }
\label{table}
\end{center}
\end{table}
The location of the hidden brane is
\begin{equation}
y_1= \ln\left({{\phi_1 \over \phi_0}}\right)^{1 \over 2}\,.
\label{locations}
\end{equation}
The electroweak scale is 
\begin{equation}
M_{EW} \simeq M_X\,e^{A(0)}=M_X\,e^{-{\phi_0^2 \over 24}}\, .
\end{equation}
The Planck scale is given by
\begin{equation}
 \left(2\,{M_{Pl} \over M_X}\right)^2 =  
\left({\phi_0^2 \over 12}\right)^{-{\xi \over 12}}\, 
\Gamma \left({\xi \over 12},{\phi_1^2 \over 12},\infty\right)
+ \left({\phi_0^2 \over 12}\right)^{-{\eta \over 12}}\,
\left[\Gamma\left({\eta \over 12},
{\phi_0^2 \over 12},{\phi_1^2 \over 12}\right)+\,
\Gamma\left({\eta \over 12},{\phi_0^2 \over 12},
{\infty}\right)\right]\,,
\label{Planck}
\end{equation}
where the generalized incomplete gamma function is 
\mbox{$\Gamma (a,x,y) \equiv \int_{x}^{y} t^{a-1}\, e^{-t}\, {dt}\,$}.
Consistency conditions imposed by positivity of the tension of the
observable brane and the profile of the radion are
$\eta<\phi_0^2<\phi_1^2\,$.
When the correct hierarchy is generated, by far 
the dominant contribution to $M_{Pl}$ comes from the first 
term on the right-hand-side of Eq.~(\ref{Planck}). 
This term is the integral over the space $y>y_1$. The condition under which
this integral dominates is $\phi_0^2<\phi_1^2<\xi$. 
Then $\eta<\xi\,$, and the brane at $y_1$ has negative tension. 
If we choose $\eta>\xi\,$, the brane will have positive tension, but
 the desired hierarchy of scales cannot be obtained.
It is not possible to place the negative 
tension brane at an orbifold fixed point because the space beyond 
$y_1$ is crucial for generating the scale hierarchy.
 We can again obtain a
solution with a string scale anywhere between $M_{EW}$ and $M_{Pl}$.

As an explicit realization that solves the
hierarchy problem, consider the following choice of parameters:
$\eta= 12\,,\phi_0^2=24\,,\phi_1^2=100\,,\xi=450$. 
The largest hierarchy among these parameters is only ${\cal O}(10)$.
With the above choice,
$M_{Pl} \approx 10^{15}\, M_X\,$ and \mbox{$M_{EW} \approx 10^{-1}\, M_X\,$}.  

{\bf{VI. Conclusion.}} 
We have studied the cosmology and hierarchy 
in models with branes immersed in 
a five-dimensional curved spacetime subject to radion stabilization. 
We found that when the radion field is time-independent and the inter-brane
spacing is stabilized, 
consistent solutions that reproduce the conventional cosmological equations 
can naturally lead to a radiation-dominated universe. 
This feature is independent of the form of the stabilizing potential.
The only assumption made is that the warp factor is symmetric 
on either side of the observable brane.

Guided by constraints on the stabilizing superpotential 
imposed by cosmology, we proceeded
to consider solutions to the hierarchy problem. We insisted that the
observable brane have positive tension and considered a noncompact
fifth dimension. We examined two classes of
models--- an exponential and a polynomial superpotential. We find 
that these scenarios generically require at least one hidden brane with 
negative tension to get the correct
hierarchy. This brane cannot be located at the
fixed point of an orbifold. 
In both models, the correct hierarchy between the electroweak 
and Planck scales can be obtained for any value of the string 
scale, including the 
interesting result $M_X \simeq M_{EW}$, without fine-tuning. The exponential 
superpotential leads to the interesting case of a vanishing
bulk cosmological constant, referred to as a self-tuning brane model. 
 As in  \cite{cosmoprob}, we needed to
truncate the space to avoid curvature singularities. 
In this model, generating the hierarchy of scales results in 
the brane tensions becoming hierarchical.  
 In the case of a polynomial superpotential no new hierarchy is created.

{\it{Acknowledgments.}} 
We thank C.~Goebel for discussions.
This work was supported in part by a DOE grant No. DE-FG02-95ER40896, 
in part by the Wisconsin Alumni Research Foundation,
and in part by the Fermi National Accelerator Laboratory, 
which is operated by the Universities research
Association, Inc., under contract No.~DE-AC02-76CHO3000.
\\

\end{document}